\documentclass[jkps,preprint,fleqn,showkeys]{revtex4}
\usepackage{graphicx}
\usepackage{amssymb}
\usepackage{amsmath}
\usepackage{bm}
\begin{document}
\setcounter{page}{0}
\title[]{Geant4 simulation model of electromagnetic processes in oriented crystals for accelerator physics}

\author{Alexei \surname{Sytov}}
\affiliation{INFN Ferrara, via Saragat 1, 44122 Ferrara, Italy}
\affiliation{Korea Institute of Science and Technology Information (KISTI), 245 Daehak-ro, Yuseong-gu, Daejeon, 34141, Korea}

\author{Laura \surname{Bandiera}}
\affiliation{INFN Ferrara, via Saragat 1, 44122 Ferrara, Italy}

\author{Kihyeon \surname{Cho}}
\email{cho@kisti.re.kr}
\affiliation{Korea Institute of Science and Technology Information (KISTI), 245 Daehak-ro, Yuseong-gu, Daejeon, 34141, Korea}

\author{Giuseppe Antonio Pablo \surname{Cirrone}}
\affiliation{National Institute for Nuclear Physics (INFN), Laboratori Nazionali del Sud, Via Santa Sofia 62, 95123 Catania, Italy}

\author{Susanna \surname{Guatelli}}
\affiliation{Centre for Medical Radiation Physics, University of Wollongong, Wollongong NSW 2522, Australia}

\author{Viktar \surname{Haurylavets}}
\affiliation{Institute For Nuclear Problems, Belarusian State University, Bobruiskaya 11, 220030 Minsk, Belarus}

\author{Soonwook \surname{Hwang}}
\affiliation{Korea Institute of Science and Technology Information (KISTI), 245 Daehak-ro, Yuseong-gu, Daejeon, 34141, Korea}

\author{Vladimir \surname{Ivanchenko}}
\affiliation{European Organization for Nuclear Research (CERN), 1211 Geneva 23 Switzerland}

\author{Luciano \surname{Pandola}}
\affiliation{National Institute for Nuclear Physics (INFN), Laboratori Nazionali del Sud, Via Santa Sofia 62, 95123 Catania, Italy}

\author{Anatoly \surname{Rosenfeld}}
\affiliation{Centre for Medical Radiation Physics, University of Wollongong, Wollongong NSW 2522, Australia}

\author{Victor \surname{Tikhomirov}}
\affiliation{Institute For Nuclear Problems, Belarusian State University, Bobruiskaya 11, 220030 Minsk, Belarus}

\date[]{Received }

\begin{abstract}
Electromagnetic processes of charged particles interaction with oriented crystals provide a wide variety of innovative applications such as beam steering, crystal-based extraction/collimation of leptons and hadrons in an accelerator, a fixed-target experiment on magnetic and electric dipole moment measurement, X-ray and gamma radiation source for radiotherapy and nuclear physics and a positron source for lepton and muon colliders, a compact crystalline calorimeter as well as plasma  acceleration in the crystal media. One of the main challenges is to develop an up-to-date, universal and fast simulation tool to simulate these applications. 

We present a new simulation model of electromagnetic processes in oriented crystals implemented into Geant4, which is a toolkit for the simulation of the passage of particles through matter. We validate the model with the experimental data as well as discuss the advantages and perspectives of this model for the applications of oriented crystals mentioned above.
\end{abstract}


\maketitle

\section{INTRODUCTION}

Crystalline structure is a unique environment for strong-field QED effects involving both high-energy charged particles and photons. Very strong electric fields of atomic planes or axes reaching up to TeV/cm drastically enhance the processes of particle deflection, radiation and electron-positron pair production. Consequently, a small piece of crystal material could be useful for a wide range of applications in accelerator physics, high-energy frontier physics, nuclear physics, cosmic-ray detectors, dark matter search and radiation therapy.

In particular, a bent crystal is able to deflect charged particle beam. This has already found its application for beam collimation and extraction at such high-energy hadron machines as U70, SPS, RHIC, Tevatron and LHC \cite{U70,RHIC,Tevatron,Tevatron2,SPSUggerhoj,UA9,UA92,LHC,LHCion}. Since such kind of deflection is equivalent to motion in a magnetic field exceeding 100 T, a bent crystal can be also used for particle spin rotation at an ultrashort distance. Therefore, one can measure a magnetic dipole moment (MDM) and electric dipole moment (EDM) of exotic fast-decaying baryons and $\tau$-leptons at high energy machines such as the LHC \cite{Baryshevsky,Baryshevsky2,MDMold,LHCMDM,LHCMDMbagli,PRD} to search for a physics beyond the Standard Model.

Due to oscillatory-like motion in a crystal structure, a charged particle, i.e. $e^\pm$ produces hard X-ray and $\gamma$ radiation \cite{ChannelingRadiation1,ChannelingRadiation2,CB0,CB}. This radiation source can be applied in nuclear physics experiments and for radiation therapy as well. Hard $\gamma$-rays can be also converted into electron-positron pair which is suitable for an intense crystal-based positron source \cite{PositronSource,PositronSource2} for both linear and circular $e^+e^-$ colliders, such as International Linear Collider (ILC), Future Circular Collider (FCC-ee) as well as for muon colliders. Crystal structure is also capable to enhance the process of pair production by hard gamma-ray which together with intense radiation leads to acceleration of electromagnetic shower development \cite{PRL2018, NIMA}. This allows one to design a compact crystalline electromagnetic calorimeter, which can be applied both in accelerator experiments and at a gamma-ray space telescope \cite{PRL2018,NIMA, Calorimeters}.

Being almost transparent for charged particles in the direction of crystal planes and axes, a crystal can be used as a media for ultra-high gradient plasma acceleration \cite{Tajima,E336}.

The design of each of these applications requires complicated simulations of entire experimental setup, usually involving High Performance Computing (HPC) on a supercomputer. One of the most suitable simulation packages for this purpose is the Geant4 simulation toolkit \cite{Geant4,Geant42}. It contains a very wide range of simulation models of particles interaction with matter and is applied in high energy, nuclear and accelerator physics, medical and space science. However, most of physics in crystals as well as of the applications mentioned above are still missing in Geant4.

In this paper, we present the new Geant4 simulation model of electromagnetic processes in oriented crystals. We validate it with the experimental data as well as we discuss its perspectives in all the applications mentioned above.

\subsection{Basics of the electromagnetic processes in oriented crystals}

\begin{figure}
\includegraphics[width=8.0cm]{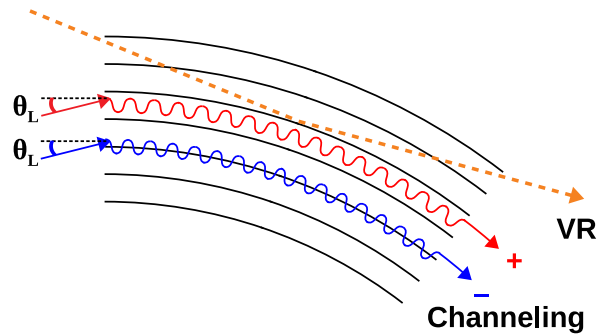}
\caption{(Color online) Schematic representation of channeling and volume reflection effects. ``$+$'' and ``$-$'' indicate channeling trajectories of positively and negatively charged particles respectively. $\theta_L$ indicates the critical channeling angle. Black lines indicate crystal planes.}
\label{Fig1}
\end{figure}

A charged particle directed almost parallel to a crystal axis/plane will follow this axis/plane, being confined by interplanar/interaxial electric field and consequently experiencing oscillatory-like motion, so-called \textit{channeling} effect \cite{Lindhard}. This motion is naturally accompanied by \textit{channeling radiation} (CR) \cite{ChannelingRadiation1,ChannelingRadiation2}. This effect is possible both in a straight and in a \textit{bent crystal} \cite{Tsyganov} as shown in Figure \ref{Fig1}. One can also notice that the positively charged particles oscillate between the crystal planes/axes while the negatively charged ones cross them every oscillation. This leads to a considerable decrease of the steering efficiency for negative particles in comparison with the positive ones due to an increase of multiple scattering, which knocks out the particles from the channeling mode, so-called \textit{dechanneling} process \cite{dechanneling,dechanneling2}.

The angle of deflection under the channeling conditions is nearly equal to the angle of crystal bending. The maximal particle-to-plane/particle-to-axis angle, at which channeling is still possible, is called the \textit{critical channeling angle} or the \textit{Lindhard angle}:

\begin{equation}
{\theta_{L} = \sqrt{\frac{2U_0}{pv}}},
\label{Eq1}
\end{equation}
where $p$ and $v$ denote the particle momentum and the velocity, respectively, $U_0$ the depth of the potential well created by the interplanar/interaxial electric field. This angle varies from $\sim$ mrad at MeV energies down to $\sim$ $\mu$rad at TeV energies, depending also on the material, the alignment and the bending angle of the crystal.

If a particle-to-plane/axis incidence angle slightly exceeds $\theta_L$, the influence of crystalline structure on the particle motion is still preserved. For instance, a charged particle of either sign can be also reflected by a bent crystal plane, so-called \textit{volume reflection} (VR) \cite{VR} (Figure \ref{Fig1}) also accompanied by radiation \cite{VRRad,VRRad2}. The angle of this reflection is comparable with $\theta_L$ and is usually considerably less than the deflection under the channeling conditions.

At higher beam-to-plane/axis angles up to few mrad it may still experience the influence of crystalline lattice, enhancing the bremsstrahlung process, called \textit{coherent bremsstrahlung} (CB) \cite{CB0,CB}. A similar effect exists for gamma propagating in a crystal, namely \textit{coherent $e^\pm$ pair production} (CPP) \cite{CB0,CB}. 

\section{Applications}

The electromagnetic processes in oriented crystals were observed experimentally in a wide energy range from MeV up to TeV and found a lot of applications. Some of these applications, both already in use and hypothetical are discussed below.

\subsection{Crystal-based extraction/collimation}

\begin{figure}
\includegraphics[width=8cm]{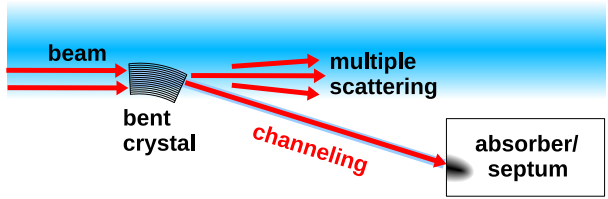}
\caption{ (Color online) Schematic representation of crystal-based collimation and extraction \cite{DESYII}.}
\label{Fig2}
\end{figure}

The idea of crystal-assisted beam steering, firstly proposed by Tsyganov in 1976 \cite{Tsyganov}, relies on planar channeling in a bent crystal.

Crystal-based beam collimation and extraction has been already applied at several proton and heavy ion synchrotrons, such as U70, SPS, RHIC, Tevatron and LHC \cite{U70,RHIC,Tevatron,Tevatron2,SPSUggerhoj,UA9,UA92,LHC,LHCion}. The main concept consists in interception of beam halo by a bent crystal and consequent deflection under the channeling conditions as illustrated in Figure \ref{Fig2}. The only difference between collimation and extraction is absorption by collimators or extraction of a deflected beam using a septum-magnet, respectively. As an advantage, this technique implies a parasitic mode without influence on the main beam fraction and on the main accelerator experiments as well. Manufacturing and installation of a crystal is much cheaper than of any other deflector, while extraction efficiency may approach 100\% for positively charged particles. 

Moreover, the same technique of extraction was proposed for electrons at few GeV energies being typical for tens of synchrotron light sources existing in the world \cite{DESYII}. The main limitation is significantly lower steering efficiency under the channeling conditions than for positively charged particles. However, the recent experiments at both Mainz Mikrotron MAMI (855 MeV $e^{-}$) \cite{PRL2014,EPJC2017} and SLAC FACET facility (multi-GeV $e^{-}$) \cite{SLAC2014,PRL2017} with the new generation of bent crystals developed demonstrated the steering efficiency in the range from 10 to 40\%, i.e. of the same order of magnitude as for protons.

Crystal-based collimation is an efficient way to protect superconducting magnets of high-energy colliders from too high radiation load. Extracted primary charged particle beams, including electron beam are of interest for testing of particle and nuclear physics detectors and generic detector R\&D \cite{fromDESYpaper}. The beams extracted from the colliders will help to carry out unique fixed-target experiments at extremely high energies.

\subsection{Magnetic and electric dipole moment measurement}

\begin{figure}
\includegraphics[width=10cm]{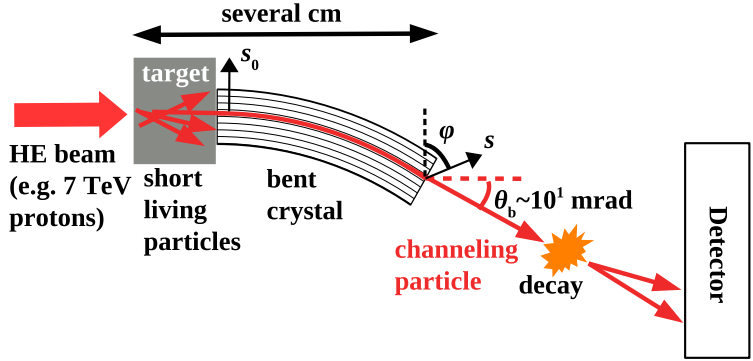}
\caption{ (Color online) Schematic representation of an experiment on MDM and EDM measurement. High energy beam interacts with a target producing exotic particles. Some of them will be deflected by a bent crystal under the channeling conditions before their decay. The decay products will be measured by the dedicated detector, which will provide the angle of spin rotation $\varphi$.}
\label{Fig3}
\end{figure}

One of the possible experiments with the extracted beam at the LHC can be an experiment on the measurement of MDM and EDM of charm baryons or tau-leptons by means of a bent crystal as illustrated in Figure \ref{Fig3}. This technique has been already tested at Fermilab for 375 GeV/c $\Sigma^+$ hyperons produced by 800 GeV/c proton beam onto a Cu target \cite{MDMold}. The MDM measured in this proof-of-principle experiment was consistent with the results of conventional measurements.

A possible dedicated experiment at the LHC \cite{LHCMDM,LHCMDMbagli,PRD} will be supplied by a $\sim$7 TeV proton beam extracted using the crystal-based extraction technique described above. This beam will be directed onto a tungsten target to produce charm baryons, i.e. $\Lambda_c^+$, $\varXi_c^+$ which will be immediately deflected by a bent crystal attached to this target. Their total thickness is chosen accordingly with the particle decay distance (typically few cm at TeV energy) to prevent a decay before the end of the crystal. The typical value of the bending angle is $\theta_b\sim10^1$ mrad, which should be also optimized to maximize the experiment sensitivity. The angle of the spin rotation $\varphi$ will be measured by measuring the kinematics of the decay products using a dedicated detector system. The spin rotation plane depends on the dipole moment type, either MDM or EDM. This measurement will provide the MDM and possibly the EDM values for exotic baryons, which cannot be measured by any other known method due to very low particle lifetime. This technique is a great opportunity to check the Standard Model predictions as well as the theories beyond the Standard Model \cite{Baryshevsky,Baryshevsky2}.

\subsection{Crystalline radiation source}

\begin{figure}
\includegraphics[width=8cm]{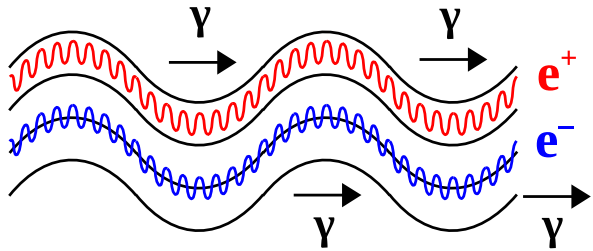}
\caption{ (Color online) Schematic representation of the crystalline undulator. ``$+$'' and ``$-$'' indicate channeling trajectories of positively and negatively charged particles respectively. Black lines indicate crystal planes.}
\label{Fig4}
\end{figure}

Since the radiation produced either in straight or bent crystals (CR and CB) is much more intense than bremsstrahlung, crystals become very attractive as innovative sources of intense coherent hard X- and $\gamma$-ray source \cite{PRL2013,PRL2015,MAMIRAD,sol,sol11}. One of the very perspective applications of radiation source is a periodically bent crystal, so-called crystalline undulator (CU) \cite{CU1,CU2,CU3,sol,sol11,sol2,CU4}, illustrated in Figure \ref{Fig4} both for $e^+$ and $e^-$.  Generally, due to lower multiple scattering angle, positron beams are more preferable than the electron ones, though they are considerably less accessible.

The idea of CU is analogical to undulator but using crystalline atomic fields instead of periodic dipole magnets. Due to high intensity of atomic fields one may generate multi-MeV gamma-rays instead of X-rays in contrast to canonical undulator, while the brilliance remains comparable to conventional X-ray sources (see the work of MBN research team \cite{sol,sol11,CU3,sol2}). This makes such kind of radiation source to be very attractive for nuclear spectroscopy, nuclear transmutation and radiation therapy.

Hard $\gamma$-ray at $\sim$GeV energy needed for accelerator physics and high energy physics can be produced in the axial field of heavy material crystals, such as tungsten. This type of radiation source can be used for positron source.

\subsection{Hybrid crystal-based positron source}

A conventional positron source for a positron accelerator is based on conversion of several-GeV $e^-$ into $e^+$ in a heavy material target, e.g. tungsten. The main limitation is the Peak Energy Deposition Density (PEDD) in this target, that can naturally reach the critical values for the case of $e^+e^-$ collider projects. 

\begin{figure}
\includegraphics[width=8cm]{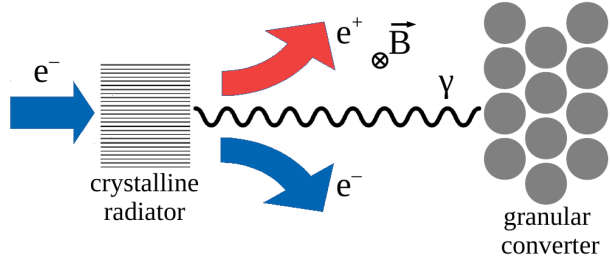}
\caption{ (Color online) Schematic representation of a hybrid crystal-based positron source.}
\label{Fig5}
\end{figure}

A crystal-based hybrid positron source may become a smart solution of this problem \cite{PositronSource,PositronSource2}. The basic idea is to split the positron converter into a $\gamma$-quanta radiator and gamma-to-positron converter as shown in Figure \ref{Fig5}. A multi-GeV $e^-$ beam passed through a thin oriented crystal will radiate an intense $\gamma$ beam. This beam and $e^\pm$ as well will diverge while passing some distance $\sim$m before the converter target. The beam spot at the converter is much wider than in the conventional scheme which drastically reduces PEDD. In some schemes it can be additionally reduced by the application of the magnetic field between the crystal and the converter target to extract all the particles except photons. A granular converter target made of small spheres can make the thermal dissipation easier. Since the positrons produced possess mostly considerably lower energies than a the primary electron beam (typically $\sim10^1\div10^2$ MeV), a large transverse size of the beam at the converter exit will not cause any significant reduction of the efficiency of the positron capture system.

A preliminary variant of a crystal-based positron source using the same oriented crystal both as radiator and a converter has been already tested experimentally and its efficiency has been demonstrated at CERN and KEK \cite{OldPositronSource1,OldPositronSource2}. Hybrid positron source for FCC-ee is currently under development \cite{PositronSource2}. This technique can be also useful for ILC and muon collider \cite{LEMMA} projects.

\subsection{Compact crystalline e.m. calorimeter}

\begin{figure}
\includegraphics[width=4cm]{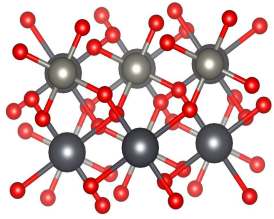}
\caption{ (Color online) The structure of the crystal lattice of a PWO scintillator crystal that can be used as a compact crystalline e.m. calorimeter.}
\label{Fig6}
\end{figure}

The idea of a compact crystalline electromagnetic calorimeter \cite{PRL2018,NIMA,Calorimeters} is based on acceleration of the electromagnetic shower in an oriented scintillator crystal due to CR, CB and CPP effects caused by the interaction of a charged particle with the field of crystal axes. The calorimeter thickness needed to absorb the energy of a high-energy particle can be reduced with a factor up to 5 in comparison with an amorphous material or a randomly oriented crystal. The lead tungstate scintillator crystal PWO, the lattice of which is shown in Figure \ref{Fig6}, and other scintillator materials usually used for conventional calorimeters, if oriented, are good condidates for a compact calorimeter. The particle-to-axis angular range of the manifestation of this process reaches several mrad, while beyond this alignment a shower is developed like in a randomly oriented crystal. Therefore, an oriented crystal can be used both as a compact calorimeter in a certain angular range and as a conventional one beyond this range.

A compact crystalline e.m. calorimeter is a smart solution for gamma-ray telescopes opening a wide prospects for $\gamma$-ray astronomy, namely to considerably reduce the dimension of satellite as well as to make available for observation sub-TeV $\gamma$ spectra \cite{Elsener}. Such kind of calorimeters may be also used in fixed-target experiment, for instance, CERN NA62/KLEVER \cite{KLEVER}, the scopes of which are to measure the branching ration of $K\rightarrow\pi \nu \nu^-$ decay modes as well as dark photon search.

\subsection{Plasma wakefield acceleration}

Plasma wakefield acceleration \cite{Tajima0} technique is a very challenging topic in accelerator physics that can be applied both in scientific experiments and industry.

\begin{figure}
\includegraphics[width=6cm]{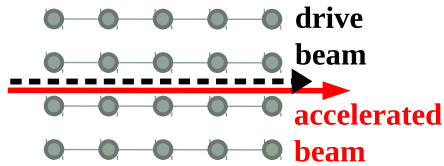}
\caption{ (Color online) Schematic representation of the plasma wakefield acceleration under the channeling conditions.}
\label{Fig7}
\end{figure}

Since solid state media possess much higher density of electrons $n_0$ than it is in gaseous plasma, its acceleration gradient $E\sim \sqrt{n_0}$ will be also much higher reaching 1 TeV/m \cite{Tajima,E336}. At the same time a crystal is almost transparent for charged particles moving under the channeling conditions. In addition, the particles are confined by the interplanar/interaxial electric field maintaining the transverse emittance for a long distance similarly to FODO elements in an accelerator. All of this makes plasma wakefield acceleration in a crystal under the channeling conditions, illustrated in Figure \ref{Fig7}, a very perspective solution to build an innovative high-energy accelerator/collider. One of the most suitable projects for this technique is a muon collider due to very tough requirements on acceleration gradients because of low muon lifetime. However, it may be also useful for many industrial applications, in particular, for medical physics \cite{E336}.

\section{New Geant4 channeling model}

Geant4 is a toolkit used worldwide for the simulation of the passage of particles through matter, including also a rich collection of electromagnetic (EM) effects of scattering and radiation. However, the channeling model existing as a Geant4 process \textbf{G4Channeling} \cite{G4Channeling} covers only the applications of crystal-based extraction/collimation and MDM, EDM measurement \cite{LHCMDMbagli} and only for hadrons. It includes neither channeling physics for electrons and positrons nor radiation and pair production in oriented crystals, which is necessary for all the remaining applications mentioned above and for $e^+e^-$ crystal-based extraction as well.

The main challenge of channeling simulations is the dependence of the value of interplanar/interaxial electric field and multiple scattering cross-section as well on the particle position in a crystal. Moreover, the radiation and pair production processes become completely different in comparison with amorphous matter. All of this requires the modification/replacement of the standard Geant4 physics processes by new ones when a particle propagates in an oriented crystal.

The currently existing realization of \textbf{G4Channeling} makes use of \textbf{G4biasing} in order to change the cross-section of standard Geant4 processes, including single Coulomb scattering, at each Geant4 simulation step. Unfortunately, this functionality does not allow one to modify so-called continuous discrete processes, including multiple scattering, ionization losses, bremsstrahlung and pair production. Though in \textbf{G4Channeling} this limitation it has a small influence on high-energy hadrons under the channeling conditions, it is crucial for $e^\pm$.

In order to switch off the standard Geant4 electromagnetic processes entirely inside the crystal volume, a Geant4 Fast Simulation Interface \cite{FastSimInterface} was chosen as a baseline for the new Geant4 Channeling model \textbf{ChannelingFastSimModel}. The Fast Simulation Interface provides a  
\textbf{G4VFastSimulationModel} abstract base model that can be inherited by a model developed. This model is active only in a certain \textbf{G4Region} defined in the geometry, only for certain particles declared in the \textbf{IsApplicable} function and only for certain conditions set up in the \textbf{ModelTrigger} function, while all the standard Geant4 physics is suspended at the same Geant4 simulation step. This fully replaces the standard Geant4 physics by the physics implemented in the \textbf{ChannelingFastSimModel} inside a chosen volume and for chosen conditions, but uses the standard physics when these conditions are not fulfilled. Therefore, \textbf{ChannelingFastSimModel} is independent of the Geant4 Physics List and, consequently, is simply transferable to the Geant4 examples already existing.

The entire replacement of the physics requires the introduction of the new physical processes into \textbf{ChannelingFastSimModel} depending on the application. All of the applications require the implementation of a \textit{basic channeling model} including an electric field function, a multiple and single Coulomb scattering model in an oriented crystal and a geometry of crystal planes/axes either bent or straight inside a crystal volume. All of these features were successfully implemented into \textbf{ChannelingFastSimModel} using the algorithms of the CRYSTALRAD simulation code \cite{CRYSTALRAD} as a baseline. In addition, a model of Coulomb scattering in an oriented crystal well validated experimentally \cite{CSS} has been also introduced. This model takes into account the division of the Coulomb scattering process onto a coherent part describing motion in the field of the crystal lattice and an incoherent part, i.e. scattering on single atoms. This model of scattering is a fundamental difference between \textbf{ChannelingFastSimModel} and \textbf{G4Channeling} and is essential for correct simulations of channeling $\sim$ GeV $e^\pm$ and potentially improves the accuracy of the simulation of high energy channeling hadrons as well. All differences are summarized in Table \ref{table1}.

\begin{figure}
\includegraphics[width=10cm]{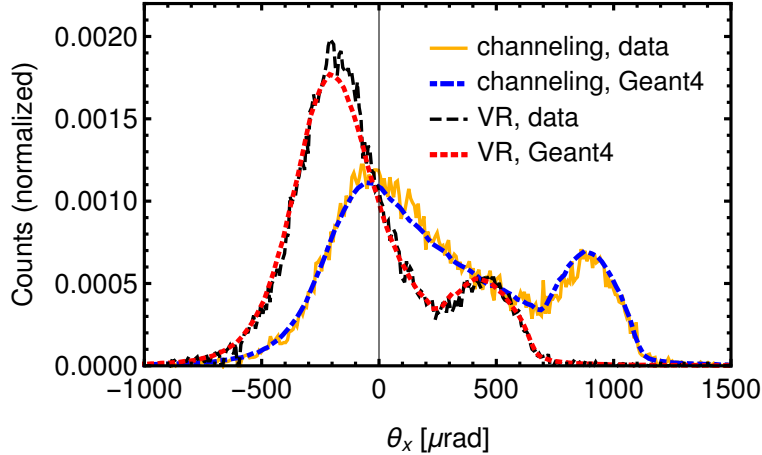}
\caption{ (Color online) Experimental \cite{PRL2014} and simulated with Geant4 \textbf{ChannelingFastSimModel} angular distributions of deflected 855 MeV $e^-$ beam by 30.5 $\mu$m thick silicon (111) bent crystal planes with a bending angle of 905 $\mu$rad for both planar channeling and VR cases. The crystal was aligned at 0 and 450 $\mu rad$ far from channeling orientation, respectively.}
\label{Fig8}
\end{figure}

\begin{table}
\caption{The main differences between \textbf{G4Channeling} and \textbf{ChannelingFastSimModel}.}
\begin{ruledtabular}
\begin{tabular}{cccc}
model & particles & Coulomb scattering & limitations \\
\colrule
 \textbf{G4Channeling} & hadrons& modified Geant4 single & continuous discrete\\&&scattering cross-section& processes remain standard \\
 \colrule
 \textbf{ChannelingFastSimModel} & $e^\pm$ &a special model of Coulomb & standard physics is \\ &&scattering in an oriented crystal&  switched off\\
\end{tabular}
\end{ruledtabular}
\label{table1}
\end{table}

The experimental validation of \textbf{ChannelingFastSimModel} was carried out using the experimental data published in \cite{PRL2014} on 855 MeV $e^-$ deflection by 30.5 $\mu$m thick silicon (111) (non-equidistant) bent crystal planes with a bending angle of 905 $\mu$rad for both planar channeling and VR cases. The crystal was aligned at 0 and 450 $\mu rad$ far from channeling orientation, respectively. The model particle-to-plane angular range is set up in \textbf{ModelTrigger} as $\pm 100 \theta_L$ to ensure the activation of \textbf{ChannelingFastSimModel} for all important angles. The comparison of the simulations and experimental data is shown in Figure \ref{Fig8} manifesting a very good agreement. The peak on the right corresponds to the beam fraction deflected under the channeling conditions while the left one includes volume reflection and multiple scattering processes. It is important to underline that the radiation energy losses, being the main process of energy losses of $e^-$ at this energy and higher, are almost negligible in this case since the crystal is very short. Therefore, \textbf{ChannelingFastSimModel} is completely adequate at these conditions. 

\begin{figure}
\includegraphics[width=10cm]{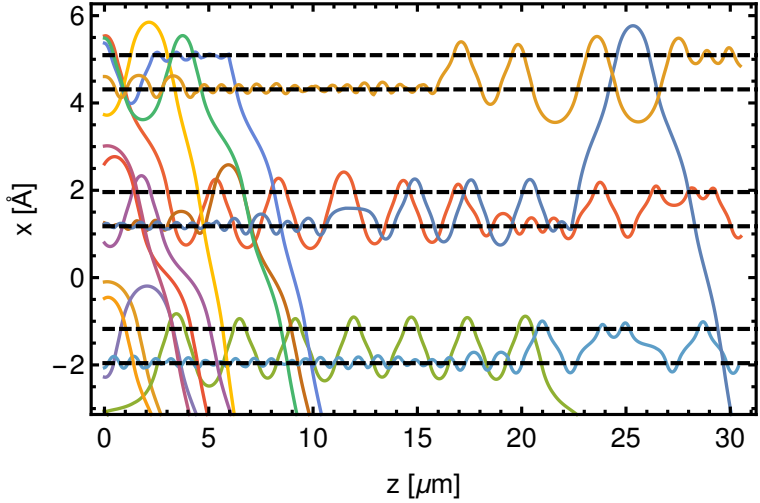}
\caption{ (Color online) Examples of $e^-$ trajectories in co-rotating reference system generated with \textbf{ChannelingFastSimModel} for planar channeling case for the same conditions as in Figure \ref{Fig8}. Horizontal dashed lines represent (111) Si crystal planes.}
\label{Fig9}
\end{figure}

The examples of $e^-$ trajectories in co-rotating reference system (with coordinate $z$ directed along and $x$ orthogonally to bent crystal planes) generated using \textbf{ChannelingFastSimModel} for the case of channeling orientation are illustrated in Figure \ref{Fig9}. One can observe that only few particles stay in channeling by the end of the crystal while most of them are kicked out by multiple scattering.

The estimated low energy limit of the model applicability is $\sim$100 MeV to ensure the quantum effects become negligible \cite{Quantum1, Quantum2}. The high energy limit is defined by the level of radiation energy losses which increases with the energy and depends on the crystal thickness and geometry. In many channeling experiments carried out for channeling of $e^\pm$ in a Si bent crystal in the energy range 0.1-20 GeV \cite{PRL2014,EPJC2017,SLAC2014,PRL2017} this model is applicable.

However, different applications mentioned above may require other simulation models and/or extensions of \textbf{ChannelingFastSimModel}, namely a \textit{radiation model} of energy losses by charged particles and of photon production, a \textit{pair production model} (by photons), \textit{deposited energy scoring}, an \textit{ionization model} and an \textit{hadron scattering model} for channeling hadrons etc.
The list of the requirements depending on the application is the following:
\begin{enumerate}

\item \textbf{Crystal-based extraction/collimation}: the \textit{basic channeling model}, a \textit{radiation model} for $e^\pm$ in the case when the radiation losses become important, an \textit{ionization model} and an \textit{hadron scattering model} for hadrons except the case when a crystal is short enough to neglect these processes, a \textit{complicated geometry model} to describe the case when many particles enter/exit the crystal through its lateral surface (usually when the beam emittance is low enough) \cite{miscut}. This application also requires the simulation of \textit{accelerator dynamics} using either existing tracking codes or by introducing a new model into Geant4.

\item \textbf{MDM and EDM measurement}: the \textit{basic channeling model}, an \textit{ionization model} and an \textit{hadron scattering model} for exotic baryons, a \textit{spin rotation model} using BMT-equations \cite{Quantum1}.

\item \textbf{Crystalline radiation source}: the \textit{basic channeling model}, a \textit{radiation model} and a \textit{complicated geometry model} for CU to simulate periodically bent crystal planes.

\item \textbf{Hybrid crystal-based positron source} and \textbf{compact crystalline e.m. calorimeter}: the \textit{basic channeling model}, a \textit{radiation model}, a \textit{pair production model} and \textit{deposited energy scoring}.

\item \textbf{Plasma wakefield acceleration}: the \textit{basic channeling model} and the particle-in-cell (PIC) method \cite{PIC} to simulate a complicated collective motion of the drive beam producing plasma waves and the accelerated beam as well. 

\end{enumerate}

\section{CONCLUSIONS}

Electromagnetic processes in oriented crystals can be used in a wide range of applications, namely crystal-based extraction/collimation of charged particles of either sign at modern synchrotrons, MDM and EDM measurement to test the physics beyond the Standard Model, crystalline radiation source and crystalline undulator for nuclear physics and radiation therapy, hybrid crystal-based positron source for future $e^\pm$ and muon collider projects, compact crystalline calorimeter for HEP experiments including dark photon search and for $\gamma$-ray space telescopes and plasma wakefield acceleration with extremely high acceleration gradients for a muon collider and, potentially, for many industrial applications, including medical physics.

Geant4 simulation toolkit is a tool that can allow one to design each of these applications and to carry out the simulations of an entire experimental setup. However, the existing Geant4 channeling model partially covers only crystal-based extraction/collimation and MDM and EDM measurement and only for hadrons, being non-applicable for $e^\pm$.

In this paper we have introduced a \textbf{ChannelingFastSimModel} Geant4 simulation model that is capable to provide a basic functionality for all the applications mentioned above. It uses Geant4 Fast Simulation Interface to suspend standard Geant4 processes and includes charged particle motion in an electric field in an oriented crystal, a multiple and a single Coulomb scattering mode and a geometry of crystal planes/axes either bent or straight inside of a crystal volume. Fast Simulation Interface makes this model completely independent of the Geant4 Physics List.

\textbf{ChannelingFastSimModel} has been validated with the experimental data for 855 MeV $e^-$ deflection in an ultrashort bent silicon crystal at Mainz Mikrotron MAMI showing a good agreement with the experimental data. It is applicable for $e^\pm$ motion in a crystal short enough, when the radiation energy losses can be neglected. This covers many experiments carried out in the energy range 0.1-20 GeV. This model will become a baseline for the development of simulation tools for the design of different applications of oriented crystals.

\begin{acknowledgments}
A. Sytov is supported by the European Commission (TRILLION, GA. 101032975). We acknowledge partial support of the INFN through the MC-INFN and INFN OREO projects; H2020-MSCA-RISE N-LIGHT (G.A. 872196) and EIC-PATHFINDER-OPEN TECHNO-CLS (G.A. 101046458) projects. We acknowledge the CINECA award under the ISCRA initiative, for the availability of high performance computing resources and support. This work is also supported by the Korean National Supercomputing Center with supercomputing resources including technical support (KSC-2022-CHA-0003). A. Sytov acknowledges the E336 collaboration for a fruitful discussion about the wakefield acceleration in a crystal.

\end{acknowledgments}

\end{document}